# High Tc Josephson nanoJunctions made by ion irradiation : characteristics and reproducibility.

Jérôme Lesueur, Nicolas Bergeal, Martin Sirena, Xavier Grison, Giancarlo Faini, Marco Aprili, Jean P. Contour

*Abstract*— Reproducible High $T_c$ Josephson junctions have been made in a rather simple two-step process using ion irradiation. A microbridge 1 to 5 µm wide is firstly designed by ion irradiating a c-axis-oriented $YBa_2Cu_3O_7$ film through a gold mask such as the unprotected part becomes insulating. A lower $T_c$ part is then defined within the bridge by irradiating with a much lower dose through a 20 wide narrow slit opened in a standard electronic photoresist. These planar junctions, whose settings can be finely tuned, exhibit reproducible and nearly ideal Josephson characteristics. Non hysteretic Resistively Shunted Junction (RSJ) like behavior is observed, together with sinc Fraunhofer patterns for rectangular junctions. The $I_cR_n$ product varies with temperature ; it can reach a few mV. The typical resistance ranges from 0.1 to a few ohms, and the critical current density can be as high as 30 kA/cm². The dispersion in characteristics is very low, in the 5% to 10% range. Such nanojunctions have been used to make microSQUIDs (Superconducting Quantum Interference Device) operating at Liquid Nitrogen ($LN_2$) temperature. They exhibit a very small asymmetry, a good sensitivity and a rather low noise.

The process is easily scalable to make rather complex Josephson circuits.

## I. Introduction

Regarding the potential applications of Josephson Junctions (JJ) working at temperatures well above the liquid helium one, there has been a tremendous hope that high $T_c$ Superconductors (HTSc) would provide materials for such applications. However, after two decades of effort, no fully reliable technology is available yet, to make with HTSc JJ the complex circuits needed for advanced applications like Rapid Single Flux Quantum (RSFQ) logical devices for instance [1]. Despite real progress in the ramp junction technology, and very encouraging recent results, there is still room for a new technology to develop. Following the pioneer work of Katz et al [2] and Kahlmann et al [3], we have made HTSc Josephson nanoJunctions (JnJ) using ion-irradiation to locally lower the critical temperature $T_c$ on a nanoscale region within a c-axis oriented $YBa_2Cu_3O_7$ film [4]. The physical ideas behind this process are twofold : first, it is a planar technology where the Josephson coupling occurs in the most favorable situation, namely within the basal plane of HTSc ; second, there is an external finely tunable parameter which controls the junction properties, namely the defect density, through the ion dose.

HTSc properties are very sensitive to defects : increasing disorder in the material firstly reduces the superconducting properties because of the d-wave symmetry of the order parameter, and increases the resistivity because of enhanced scattering. For high enough defect density, a superconducting to insulator transition is observed. Based on these facts, we have developed a two steps strategy to make HTSc JnJ : high dose (in the $10^{15}$ at/cm² range) ion irradiation is used to draw lines, bridges, contact pads and circuits through micron-size shadow masks : a lower dose (in the $10^{13}$ at/cm² range) is then used to locally depressed the superconducting properties within a bridge, and form Superconductor-Normal-Superconductor (SNS) Josephson junctions.

## II. High Tc Josephson Nanojunctions

A t=150 nm thick c-axis oriented $YBa_2Cu_3O_{7-\delta}$ (YBCO) film is pulsed laser deposited and in-situ covered by a 40 nm thick gold layer to insure both low contact resistances to the junction and reproducible characteristics. PMMA (polymethyl methacrylate) photoresist is then deposited and patterned to design microbridges (1 to 5 µm wide, 8 to 40 µm long). A 250 nm thick gold layer is deposited and lifted-off, such as the microbridges and the contacts remains covered. The in-situ gold layer is removed by Ar Ion Beam Etching (IBE). The sample is then ion-irradiated with 100 keV oxygen ; the $5\times10^{15}$ at/cm² fluence makes the unprotected parts insulating, therefore designing a current path underneath the gold layer including the bridges and the contact pads. No HTSc material is removed during this process. In a second step, gold above the microbridge is removed by Ar IBE through a suitable PMMA mask. Finally, photoresist is deposited all over the sample, and a narrow slit (20nm width) is opened across the microbridge which will defined the junction area. 100 keV oxygen ions are used to lower the $T_c$ in this region, with typical fluences of a few $10^{13}$ at/cm². We therefore end with a junction completely embedded in a cuprate layer, contacted with in-situ low resistance gold pads. No further annealing or heat treatment is needed to obtain the characteristics described below, as opposed to previous studies [3].

Figure one displays the resistance as a function of temperature for junctions made with different ion doses

Manuscript received August 29, 2006. This work was supported in part by the Centre National de la Recherche Scientifique (France) through a post-doctoral fellowship.

J. L. , N. B., M. S., X. G. , M. A. Authors are with UPR5 CNRS, Physique Quantique, ESPCI, 10 rue Vauquelin, 75 231 Paris Cedex 05 France (corresponding author to provide phone: +33 1 40 79 45 34; fax: + 33 1 40 79 47 44 ; e-mail: jerome.lesueur@ espci.fr).

G. F. Author, is with LPN-CNRS, route de Nozay, 91 460 Marcoussis, France.

J. P. C. Author is with UMP Thales-CNRS, route départementale 128, 91 767 Palaiseau, France.

ranging from 1.5 to $6\times10^{13}$ at/cm$^2$ within a 1μm wide bridge. Below the resistance drop of the reservoirs at 90 K, a plateau develops and a complete transition to a full zero-resistance state is observed. It corresponds to the Josephson coupling through the damaged zone, at a temperature $T_J$ which can be finely tuned by the ion fluence, since the transition width is small ($\Delta T_J$ =1-2 K). This is not the transition of the irradiated part itself $T_c'$. Based on previous studies [5] [6] and on Monte-Carlo numerical simulation using the TRIM code [7], we know $T_c'$ for the corresponding defects densities, which appears to be much lower than $T_J$. This becomes experimentally clear when observing the I-V characteristics of the junctions. The main part of Figure 2 displays I-V of $6\times10^{13}$ at/cm$^2$ irradiated junction at different temperature *below* the zero resistance state observed in R(T) curves measured with a tiny current. At high temperature, a RSJ-like behavior is observed, with its *upward* curvature (see right inset blow-up), typical of a SNS Josephson junction [8]. At low temperature, a *downward* curvature is seen, as expected for a flux-flow regime (see left inset blow-up) : the whole bridge is then superconducting. The temperature which separates the two regimes is $T_c'$ [32 (±1) K in this case]. In fact, this corresponds to the lower $T_c'$ within the bridge. Due to lateral straggling of the incoming ions through the 20 nm wide aperture, there is a distribution of defects along the junction, and therefore a distribution of $T_c'$. The inset of Figure 1 displays a TRIM simulation of the defect density (dpa stands for displacement per atom), and the corresponding "local" $T_c'$, assuming an Abrikosov-Gorkov depairing model for impurity scattering in a d-wave superconductor, which has been experimental verified [5]. For a fluence of $6\times10^{13}$ at/cm$^2$ for instance, the minimum calculated $T_c'$ is of the order of 35 K (as compared to the 32 K experimental value), but the coupling temperature is almost 50 K. This is due to the proximity effect : superconducting correlations coming from high $T_c$ regions propagate within the "normal" one and set the Josephson coupling [4]. Extended calculations have been made [9] to numerically compute $T_J$ in this case.

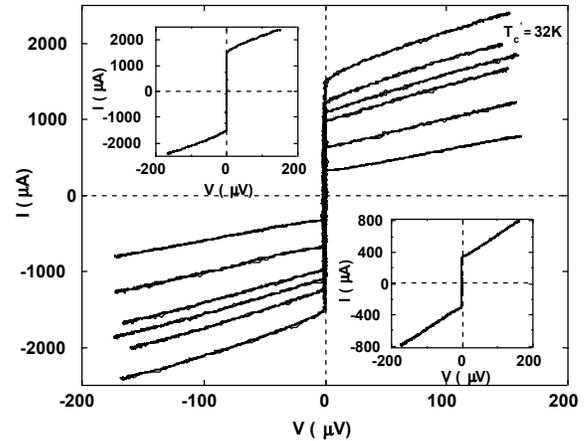

Fig. 2. I-V characteristics of a JnJ irradiated with 6 10$^{13}$ at/cm$^2$ at different temperatures (from top to bottom 29 K, 31 K, 32 K, 33K, 36 K, 40 K). At high temperature, a RSJ-like Josephson characteristic is observed, with a typical upward curvature (see inset bottom). At low temperature, the downward curvature (see inset top) indicates a flux-flow regime, as expected when the whole sample is superconducting, below $T_c'$ (here estimated to be 32 K).

The critical current $I_c(T)$ follows a quadratic dependence with the temperature whatever the irradiation dose is, as shown in Figure 3. This is consistent with the de-Gennes-Wertammer model of proximity effect for Josephson coupling [10]. Cooper pairs diffuse in the normal regions on a "normal coherence length" $\xi_N(T)$, and insure the coupling provided thermal fluctuations do not destroy phase coherence.
This can be expressed as follows :

$$\xi_N(T) = \sqrt{\frac{\hbar D}{2\pi k_B T}} \sqrt{1 + \frac{2}{\ln(T/T_c')}} \qquad (1)$$

where T is the temperature and D the diffusion constant. The critical current is therefore :

$$I_c(T) = I_0 (1 - \frac{T}{T_J})^2 \frac{l/\xi_N}{\sinh(l/\xi_N)} \qquad (2)$$

where $I_0$ is the typical critical current, $l$ the length of the normal part assuming a rectangular shape for the junction. The main temperature dependence is provided by the quadratic term, since $\xi_N(T)$ itself varies slowly with temperature [sqrt(T) and ln(T)]. It is worth noticing that the actual length of the junction $l$ is not strictly constant as a function of temperature, because of the defect (and therefore the $T_c'$) distribution within the junction. However, since it enters in a rather weak varying function [x/sinh(x) at small x], the corresponding correction to the quadratic law is small. More details about this are given in [4].

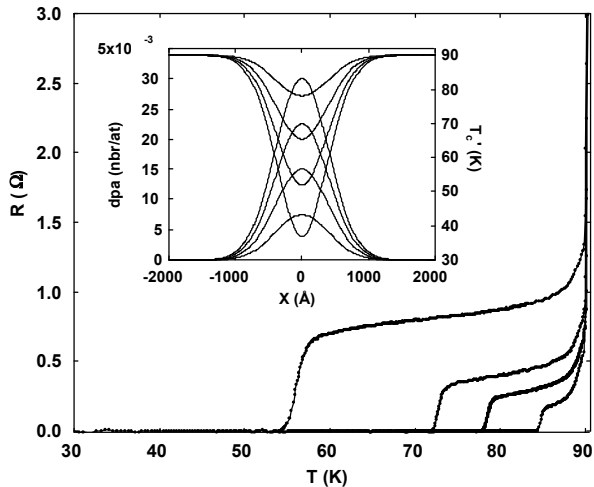

Fig. 1. Resistance as a function of temperature of 5 μm wide junctions irradiated with differente doses (1.5, 3, 4.5, $6\times10^{13}$ at/cm$^2$). The Josephson coupling temperature decreases and the resistance increases as the dose increases. Inset : computed defect density along the junction (solid line, left axis), and corresponding local critical temperature $T_c'$ (dashed line, right axis) for the same doses as the main frame (see text).

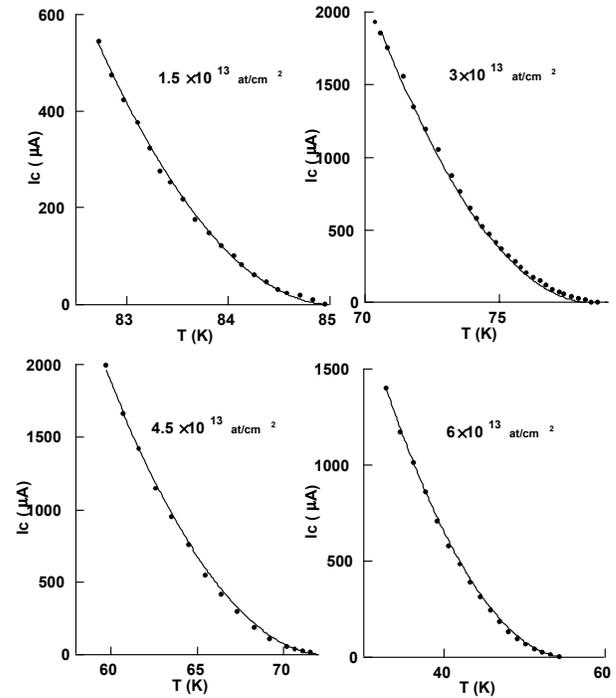

Fig. 3. Critical current as a function of temperature for JnJ irradiated with different doses. Solid lines are the quadratic fit according to de Gennes-Wertammer model (see text).

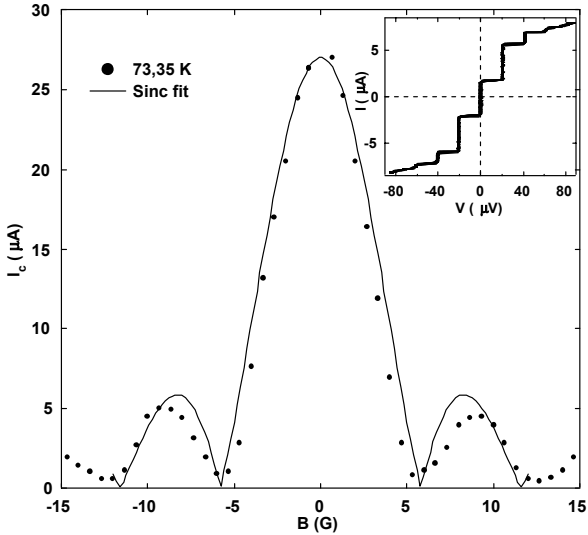

Fig. 4. Critical current as a function of the applied magnetic field for a $3\times10^{13}$ at/cm$^2$ irradiated junction measured at 73.35 K. A typical Fraunhofer pattern for a rectangular geometry is observed (Sinc fit solid line).Inset : shapiro steps observed under microwave irradiation of a $7\times10^{13}$ at/cm$^2$ irradiated junction measured at 4.2 K.

The most stringent quality test for JJ is the Fraunhofer pattern. Figure 4 displays the modulation of the critical current of a junction as a function of the applied magnetic field, and the |sinc(B)| fit expected for a rectangular junction in the small junction limit (the width of the junction $w$ smaller than the Josephson penetration length $\lambda_J$, which is the case here). The good agreement proves that the current density is very homogeneous within the junction. When the temperature is decreased, the critical current increases, and therefore $\lambda_J$ becomes smaller than $w$ : a long junction behavior is observed [4]. In addition to the dc Josephson effect, the ac one is clearly evidenced, as proved by the Shapiro steps in the I-V characteristic observed under microwave irradiation (see inset Figure 4).

The typical parameters of these junctions make them suitable for numerous applications. They are self-shunted with non hysteretic behavior. The resistance ranges from 0.1 Ω to a few Ω depending on the width of the junction and the ion dose, and can be finely tuned. The critical current varies with temperature but can easily rise up to a few mA, which corresponds to critical current densities in the 10 to 30 kA/cm$^2$. The $I_cR_n$ product is therefore in the few mV range.

The reproducibility of these characteristics is very high, especially for irradiation fluences lower than $6\times10^{13}$ at/cm$^2$. The dispersion in $T_J$ is smaller than 1 K, and that in $R_n$ in the 5 % range. Given the temperature dependence of $I_c(T)$, the dispersion in critical current varies, but remains in the 5-10% range in most practical cases. It is worthwhile mentioning that on-chip dispersion can be very low (in the 1% range for $R_n$ and $I_c$) if junctions are close to each other, typically a few 10 μm apart [9]. Systematic study of the dispersion source is on its way. The JnJ aging is negligible over months, and their characteristics do not change upon temperature cycling. These excellent performances are due to the fact that the whole device is made with the same material, therefore with a unique thermal expansion coefficient, and that the active area is embedded within the film, which minimizes the oxygen out-diffusion.

### III. HIGH TC MICROSQUIDS

We have used JnJ to make micro-SQUIDs. Two main geometries have been used for this study : SQUID 1 (S1) has a loop area of 10*10 μm$^2$ with a 5 μm arm width corresponding to an inductance L1=32pH : SQUID 2 (S2) has a loop area of 6*6 μm$^2$ and a 2 μm arm width corresponding to an inductance L2=17pH. The fabrication process is basically the same as for a single junction : the loop is made in the first step irradiation ; the junctions in the middle of the arms in the second one. Two different doses ($3\times10^{13}$ and $6\times10^{13}$ at/cm$^2$) have been used with both S1 and S2 geometries. The former dose was chosen for operation at LN$_2$ temperature. The I-V characteristics of the SQUIDs are the same as the isolated junctions.

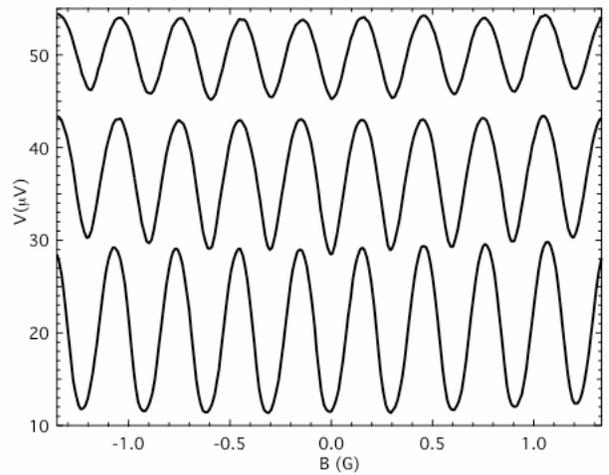

Fig. 5. Modulation of a 6 μm loop SQUID irradiated with $3\times10^{13}$ at/cm$^2$ measured at LN$_2$ temperature under different bias currents above the critical current (Ic = 15 μA). From top to bottom : 30 μA, 25 μA, 18 μA.

All the SQUIDs display periodic modulations upon magnetic field, with a period of 0.08 G (S1) and 0.3 G (S2), in good agreement with the size of the loop, provided the flux-focusing effect due to the screening of the superconducting parts is properly taken into account. As an example, Figure 5 shows the SQUID modulation of a (S2) $3\times10^{13}$ at/cm$^2$ irradiation device at LN$_2$ temperature, upon different bias currents slightly above the critical current (15 µA). As the current bias is increased, the amplitude of the oscillations decreased. The SQUID is normally operated at the bias current that gives the maximum value of the flux to voltage function transfer $V_\phi=dV/d\phi$. We obtained $V_\phi\approx55\mu V/\phi_0$ for S1 and $V_\phi\approx60\mu V/\phi_0$ for S2. At higher field, the Fraunhofer pattern of the individual junctions clearly superimposes on the SQUID cosine modulation itself. This shows that the junctions have similar characteristics, and therefore that the asymmetry of the SQUID is weak, as also proved by the direct measurement of $I_c(B)$ patterns.

Of course, noise is an important issue. We have performed preliminary noise measurements on these devices which look encouraging ($<10^{-10}$ V/$\sqrt{Hz}$ at 1 kHz) [11]. On-going experiments are being carried out in order to specify the detailed noise spectrum.

## IV. CONCLUSION

We have a developed a technology to make HTSc Josephson nanoJunctions using ion irradiation. These non-hysteretic SNS junctions have characteristics which make them suitable for numerous applications in superconducting electronics. Their resistance is low, the critical current density is in the few 10 kA/cm$^2$ range, the IcRn product can reach a few mV. All the parameters can be finely and easily adjusted by choosing the geometry and/or the ion fluence. An example has been given : a dc-SQUID operating exactly at LN$_2$ temperature. The characteristics of the JnJ are very reproducible, and can achieve impressive low on-chip dispersion : this is a key point as far as the development of complex circuits is concerned. Three main advantages of this technology have to be pointed out in the perspective of large scale applications : (i) the starting material is a standard c-axis oriented HTSc film on a *standard* substrate ; (ii) except gold for contact pads, no other material is involved, which makes the process easy and also insures a very good aging and cycling of the devices : (iii) this process is highly scalable, since there is no design constraint (the junctions can be put anywhere on the film surface) and since it is based on standard e-beam lithography methods.

Besides the SQUID, the first application we are working on is related to RSFQ logics. Following the work of Kim et al [12], we have designed a T flip-flop circuit which should operate at 30 K, requiring 6 to 8 junctions : it is in fact a dc voltage divider. Test and measurements are in progress. Another application will be Josephson arrays for voltage standard for instance, as already explored by Chen et al [13]. Superconducting THz detectors are also a major issue : Humphreys et al [14] used HTSc JJ arrays and antennas to image a room temperature scene in the 100 GHz frequency range. The HTSc JnJ presented here can be easily used to design such arrays, with on chip antennas and filters. Finally, extension of the ion irradiation technique to make JJ in superconductors has been done with MgB$_2$ successfully [15]. This may also be a interesting route for new superconducting devices.

## ACKNOWLEDGMENT

The authors gratefully acknowledge O. Kaitasov and S. Gautrot for the ion irradiation made at IRMA-CSNSM (Orsay-France), E. Jacquet, F. Lalu and L. Leroy for technical support.